\documentclass[aps,prb,twocolumn,superscriptaddress,showpacs,showkeys]{revtex4-1}
\usepackage{amsmath,amssymb}
\usepackage{graphicx,epsfig}
\usepackage{txfonts}
\usepackage{color}
\usepackage{hyperref}
\begin{document}

\title{Isotope effect in acetylene C$_2$H$_2$ and C$_2$D$_2$ rotations on Cu(001).}
\author{Yulia E. Shchadilova}
\email[]{y.shchadilova@gmail.com}
\affiliation{A. M. Prokhorov General Physics Institute, Russian Academy of Science, Moscow, Russia}
\author{Sergei G. Tikhodeev}
\affiliation{A. M. Prokhorov General Physics Institute, Russian Academy of Science, Moscow, Russia}
\affiliation{Division of Nanotechnology and New Functional Material Science, Graduate School of Science and Engineering,
University of Toyama, Toyama, 930-8555 Japan}
\author{Magnus Paulsson}
\affiliation{Department of Physics and Electrical Engineering,
Linnaeus University, 391 82 Kalmar, Sweden}
\affiliation{Division of Nanotechnology and New Functional Material Science, Graduate School of Science and Engineering,
University of Toyama, Toyama, 930-8555 Japan}
\author{Hiromu Ueba}
\affiliation{Division of Nanotechnology and New Functional Material Science, Graduate School of Science and Engineering,
University of Toyama, Toyama, 930-8555 Japan}
%\date{\today}
\date{December 26, 2013}

\begin{abstract}
A comprehensive analysis of the elementary processes behind the scanning tunneling
microscope controlled rotation of  C$_2$H$_2$ and C$_2$D$_2$,  isotopologues of a single acetylene molecule
adsorbed on the Cu(001) surface is given, with a focus on the isotope effects.
With the help of density-functional theory  we calculate  the vibrational modes
of  C$_2$H$_2$ and C$_2$D$_2$ on Cu(001) and estimate the anharmonic couplings between them,
using a simple strings-on-rods model. The probability of the elementary processes
--- non-linear and combination band --- are estimated using the Keldysh diagram technique.
This allows us to clarify the main peculiarities and the isotope effects of the C$_2$H$_2$ and C$_2$D$_2$
on Cu(001) rotation, discovered in the pioneering work  [Stipe \textit{et al.}, Phys. Rev. Lett. \textbf{81}, 1263 (1998)],
which have not been previously understood.
\end{abstract}

\maketitle
\section{Introduction}
\label{sec:Introduction}

The utmost nanotechnology, that is, a method to manipulate a single atom and/or molecule adsorbed
on a solid surface has been developed in the last twenty years as an impressive implementation of the
scanning tunneling microscope (STM) and inelastic electron tunneling spectroscopy
(IETS).\cite{Eigler1991,Mo1993,Stipe1997,Stipe1998b, Stipe1998a,Komeda2002}
 In a recent Letter~\cite{Shchadilova2013a}
we clarified the elementary processes behind one of the pioneering works on single molecule manipulation,
the rotation of a single acetylene molecule  on the Cu(001) surface.~\cite{Stipe1998a}
However, the isotope effect in the acetylene / deuterated acetylene rotations
on Cu(001) also discovered in Ref.~\onlinecite{Stipe1998a} was left beyond the scope of our letter.
The goal of the present paper is to extend our approach of Ref.~\onlinecite{Shchadilova2013a} in order to give a comprehensive
analysis of the STM-induced acetylene  C$_2$H$_2$ and deuterated acetylene C$_2$D$_2$ rotations on Cu(001),
with a special emphasis on the isotope effects.

The work of Stipe, Rezaei, and Ho~\cite{Stipe1998a}  appeared to be the first
thorough and systematic experiment on a single adsorbate manipulation
made in combination with STM-IETS. This
has been established since then as an indispensable
experimental method in order to gain insight into the vibrationally
mediated motions and reactions of single molecules with STM (see,
e.g., in Ref.~\onlinecite{Ueba2011} and references therein).

The work~\cite{Stipe1998a} was also the first where the responsible for modification vibrational mode
was not directly excited by the tunneling electrons.
The observed rotation yield $Y(V)$ per
electron as a function of bias voltage $V$ for C$_2$H(D)$_2$ exhibits the threshold
at 358~(266)~meV corresponding to excitation of the C--H(D) stretch
mode. The peak in the $\Delta{\rm log}(Y)/\Delta V$ plot agreed with
the observed IETS spectrum ($d^2I/dV^2$) for both systems.  The
threshold energy of the rotation yield $Y(V)$ and the peak in the
$\Delta{\rm log}(Y)/\Delta V$ plot indicated that a selective excitation of
the C--H(D) stretch mode is a trigger for the rotation.
In this respect, it is very different from, e.g.,  the rotation of a single oxygen molecule on  Pt(111)
surface,\cite{Stipe1998b} where the hindered rotational mode can be directly excited
by the tunneling electrons.\cite{Teillet-Billy2000}

The case of acetylene/Cu(001) with an indirect reaction pathway  is not unique,
many other examples of indirect excitation of the reaction coordinate (RC)
mode have been established, e.g., the migration of CO on Pd(110).\cite{Komeda2002}
However, the rotation of C$_2$H(D)$_2$/Cu(001) demonstrates many peculiar features.
First of all there are the lower and higher thresholds
at the bias voltage around 240--250~mV and 360~mV for the acetylene rotation,
and seemingly only a single threshold at 275~mV for the deuterated acetylene rotation.
While it is easy to assign the higher threshold to the C--H(D) stretch vibrational
mode, the lower threshold and its absence in case of   deuterated acetylene
has not been  previously discussed explicitly.
Secondly, the crossover from linear (single electron process) to nonlinear (two-electron process) dependencies
of the rotation rate of the C$_2$H$_2$/Cu(001) with increasing tunneling current has not been well understood.
This nonlinearity becomes more pronounced with increasing the bias voltage.
This is in contrast to the so-called coherent ladder
climbing where reaction rate as a function of current approaches to a linear one with the increase
of bias voltage.~\cite{Stipe1998,Salam1994}

Previously, we have shown\cite{Shchadilova2013a} that for the C$_2$H$_2$/Cu(001)
the rotational probability can be divided into three partial processes
\begin{equation}\label{eq:RABC}
R(V) = R_A(V)+R_B(V)+R_C(V),
\end{equation}
where the rates $R_A(V)$ and $R_B(V)$ are, respectively, the one- and two-electron processes with a higher threshold $V\sim 358$~mV,
and $R_C(V)$ is the one-electron combination band process, initiated via inelastic emission by tunneling electron of a pair of
acetylene on Cu(001) vibrational excitations.
We believe that such a splitting can be also done for the
deuterated acetylene on Cu(001) surface.

In this paper we analyse the linear $R_A(V)$ as well non-linear $R_B(V)$ processes for both isotopologues of
acetylene  on Cu(001). We make the throughout analysis of their vibrational modes  and estimate
the anharmonic couplings between them.
We also provide the explicit derivation of the excitation rate of the reaction coordinate mode via anharmonic
couplings and make the comparison with the experimental results of Ref.~\onlinecite{Stipe1998a}.
The analysis of the possible combination band excitations contributing to  $R_C(V)$ is also given for the both
isotopologues of acetylene on Cu(001).

The structure of this paper is the following.
In Sec.~\ref{sec:Direct_excitations} we introduce the Hamiltonian of the system and discuss the linear
process of the reaction coordinate excitation.
In Sec.~\ref{sec:DFT} we calculate the vibrational modes and lifetimes of the acetylene molecule on Cu(001) surface.
Then in Sec.~\ref{sec:Anharmonic_coeff} we estimate the anharmonic couplings between the vibrational modes.
In Sec.~\ref{sec:Keldysh_formalism} we introduce the Keldysh formalism to obtain the rate of the nonlinear process
and compare it with the experimentally obtained results in Sec.~\ref{sec:Discussion}.
The discussion of the combination band process for the acetylene isotopologues
on Cu(001) is given in Sec.~\ref{sec:Combinational}. Concluding remarks are given in Sec.~\ref{sec:Conclusions}.

\section{Model introduction. Direct excitation}
\label{sec:Direct_excitations}

In order to describe the elementary processes that occur in the acetylene molecule on Cu(001) surface we divide
the full Hamiltonian of the system on the electronic and vibrational (phonon) parts,
\begin{equation}\label{Eq:Hfull}
H = H_{e}(\lbrace \delta q_{\nu}\rbrace) + H_\mathrm{ph},
\end{equation}
where the electronic part $H_e$ depends on the normal coordinates of the molecule $\lbrace \delta q_{\nu}\rbrace$.
The electronic part of the Hamiltonian can be written in a form of the Anderson-Newns type Hamiltonian,\cite{Newns1969}
\begin{equation}
\label{eq:He}
H_e=\varepsilon_a (\lbrace \delta q_{\nu}\rbrace)c_a^{\dagger}c_a +
 \sum_{ j= t,s} \varepsilon_{j} c_{j}^{\dagger} c_{j}+
 \sum_{j= t,s} \mathcal{V}_{j} \left( c_j^{\dagger} c_a+h.c.\right),
\end{equation}
where the indices $s$($t$) and $a$ denote the substrate (tip) and adsorbate, respectively; the corresponding
energy levels are $\varepsilon_{s(t)}$ and $\varepsilon_{a}(\lbrace \delta q_{\nu}\rbrace)$.
Electronic tunneling matrix elements $\mathcal{V}_{t}$
(tip-adsorbate) and $\mathcal{V}_{s}$ (substrate-adsorbate) give rise to a stationary tunneling current between the tip
and the substrate through the adsorbate orbital at applied bias voltage $V$.
The electron occupation functions in the substrate and tip are assumed to be
Fermi  distributions with the same temperature $T$ but
different chemical potentials $\mu_t$ and $\mu_s$,
 $\mu_s-\mu_t = eV$.

The excitation of the high frequency C--H(D) stretch modes is described with the expansion of
the adsorbate orbital energy in the first term of Eq.~(\ref{eq:He})
to the first order over the electron-phonon coupling as\cite{Persson1980}
\begin{equation}
\label{eq:Ea}
\varepsilon_a(\lbrace \delta q_{h}\rbrace) \approx \varepsilon_a(0) + \chi (b_{h}^{\dagger} +b_{h}),
\end{equation}
where $\chi$ is the electron-phonon coupling constant, $ \varepsilon_a(0)$ is the unperturbed
adsorbate energy,
$b_{h}$ is the annihilation operator of the high frequency vibrational mode
$\left( \delta q_h=2^{-1/2}(b^\dagger_h+b_h) \right)$
%with the frequency $\Omega_{h}=358$~meV,
which is directly excited by the inelastic tunneling current.
%~\footnote{The energies of the high frequency C--H stretch
%modes are slightly different, see in Table~\ref{tab:VibModes}; we use the experimental value
%from Ref.\onlinecite{Stipe1998a} in the estimates below.}.

The high frequency vibration generation rate then reads~\cite{Tikhodeev2004}
\begin{equation}\label{Eq:Giet}
\Gamma_{\mathrm{iet}}(\Omega_{h},V)=\int d\omega \rho^{(h)}_\mathrm{ph}(\omega)\Gamma_{\mathrm{in}}(\omega,\Omega_{h},V),
\end{equation}
where $\rho^{(h)}_\mathrm{ph}(\omega)$ is the Lorentzian-shaped density of states
\begin{equation}
\rho^{(h)}_\mathrm{ph}(\omega)=\frac{1}{\pi}\frac{ \gamma_\mathrm{eh}^{(h)}}{
\left(\omega-\Omega_h\right)^2+ \left(\gamma_\mathrm{eh}^{(h)}\right)^2},
\end{equation}
and $ \gamma_\mathrm{eh}^{(h)}$ is the inverse lifetime of the phonon mode $h$
due to electron-hole pair excitation, specified in Tab.~\ref{tab:VibModes}.

%--------------------------------------------------------------------------------------------------------
%Table with the vibrational modes of the acetylene molecule
%--------------------------------------------------------------------------------------------------------

\begingroup
\squeezetable
\begin{table*}[t]
\caption{\label{tab:VibModes}
Calculated/experimental vibrational
energies, damping rates, and angular momentum of C$_2$H(D)$_2$ on Cu(001).
Numbers in parentheses belong to C$_2$D$_2$.
 The frustrated rotational modes along (001) are emphasized in bold.}
\begin{ruledtabular}
\begin{tabular}{c l c c c c c |c|c c c}
$\nu$
	& Mode
		& \multicolumn{5}{c}{$\hbar\Omega_{\nu}$}
			& $\gamma_\textrm{eh}^{(\nu)}$
				& $L_x^{(\nu)}$
					& $L_y^{(\nu)}$
			 			& $L_z^{(\nu)}$ \\
	&
		&  \multicolumn{5}{c}{meV}
			&      $10^{12}$s$^{-1}$
				& \multicolumn{2}{c}{(rel. u.)}        \\
\hline
	&
		&  \multicolumn{2}{c}{theory}
			&  \multicolumn{3}{c}{expt}
				&  \multicolumn{4}{c}{theory} \\
		&
			&  this
				&
					&\multicolumn{2}{c}{EELS}
							&  IETS
								&
									&
										&
											&\\
	&
		&work
			&  Ref.~\onlinecite{Olsson2002}
				& Ref.~\onlinecite{Marinova1987}
					& Ref.~\onlinecite{Avery1985}
				 		&  Ref.~\onlinecite{Stipe1998a}
				 			&  \multicolumn{4}{c}{this work} \\
\hline
1   				& C--H(D) stretch, symmetric            	& 371 (275)				& 379           	& 364          &360 (272) 			&  358             & 1.0 (0.3)		  		& -0.01 (0)			&	0.01 (0) 		&0\\
2   				& C--H(D) stretch, asymmetric            & 368 (270)             & 375           	& 357          	&360 (272) 			& 358              & 0.7 (0.6)        		& 1.6 (-1.6)			&-1.6 (1.6)		&0 \\
3   				& C--C stretch          							& 167 (164)             & 171           	& 164         	&162 (159)			&  N.O.$^{**}$ 	& 2.2 (2.4)           	& 0						& 0 					&0  \\
4    				& C--H(D) in-plane bend or wag, asymmetric
																			& 131 (108) 				& 132           	& 141          &141 (115)    		&  N.O. 			& 0.2 (0.07)      		& 1.2 (-1.2)			&-1.2 (1.2)		&0  \\
5     				& C--H(D) in-plane bend  or scissor, symmetric
																			& 111 (79)				& 117           	& 118          &117 (84)				&  N.O.				& 1.5 (0.7)   				& 0						&	0					&0  \\	
\textbf{6} 	& \textbf{C--H(D) asym rotation or out-of-plane bend}
																			&\textbf{100 (77)}	&\textbf{101} 	&\textbf{78}	&\textbf{79 (63)}	& N.O. 		&\textbf{ 0.7 (0.3)}			& 0 						& 0					&\textbf{ -1 (-1)} \\
7   				& out-of-plane bend or cartwheel   & 71 (52)   	            & 75               	& N.O.      	&N.O					& N.O. 				& 0.2 (0.1)             	& -0.2 (0.4) 			& -0.2 (0.4)		&0  \\
8    				& in-plane bend or wag
																			& 58 (50)                	& N.P.$^*$       & N.O.  		&N.O.(50)				& N.O. 				& 2.0 (1.5)             	&	1.3 (1.1)			&-1.3 (-1.1)		&0  \\
9    				& molecule-Cu stretch       				& 50 (49)               	& N.P.            	& 52				&58 (37)				& N.O.				&0.05 (0.06)            &	0 (-0.04)			&0 (0.04)			&0  \\
10    			& in-plane rotation      					& 29 (29)                	& N.P.             	&  N.O.			&N.O.					&  N.O.				&1.8 (1.7)           		&	-0.4 (-0.4)		&0.4 (0.4)			&0  \\
\textbf{11} 	& \textbf{out-of-plane rotation}		&\textbf{28 (26)}   	&  N.P.            	&  N.O.			&N.O.					& N.O.     			&\textbf{0.2 (0.2)}	&	0						&0					&\textbf{ 1 (1)} \\
12    			& out-of-plane bend     					& 23 (22)                	&  N.P.          	& N.O.			&N.O.					&  N.O. 			&0.04 (0.04)     		&	-0.04 (-0.07)	&-0.04 (-0.07)	&0.03(0.05) \\
\hline
\multicolumn{11}{l}{$^*$   N.P.: Not Published}\\
\multicolumn{11}{l}{$^{**}$  N.O.: Not Observed}\\

\end{tabular}
\end{ruledtabular}
\end{table*}
\endgroup

At $T=0$ the vibrational generation rate can be written in a simple form~\cite{Gao1997,Tikhodeev2004}
\begin{equation}
\Gamma_{\mathrm{in}}(\omega, \Omega_{h}, V)\simeq \frac{\gamma_\mathrm{eh}^{(h)}}
{\hbar \Omega_{h}} \frac{\Delta_t}{\Delta_s}\left(\left|eV\right|-\hbar\omega\right)\Theta\left(\frac{\left|eV\right|}{\hbar\omega}-1\right),
\end{equation}
where $\Theta(x)$ is the step function, and
$\Delta_s(t)=\pi \sum_{s(t)} \left| \mathcal{V}_{s(t),a}\right|^2\delta \left(\varepsilon-\varepsilon_{s(t)}\right)$
is the hybridization parameter of the tip and the adsorbate (substrate and adsorbate).

The mechanism of the energy transfer to the reaction coordinate mode
being separated on the partial processes~(\ref{eq:RABC}) can be analysed as following.
The linear partial process $R_A$ is the one-electron process of the direct excitation
over the RC barrier, it is similar to the processes described in Refs.~\onlinecite{Komeda2002,Persson2002}.
It can be written as a linear function of the C--H(D) stretch mode generation rate~\cite{Kumagai2012}
\begin{equation}
R_A(V)=A \Gamma_{\mathrm{iet}}(\Omega_{h},V) ,
\end{equation}
where the proportionality coefficient $A$ is the energy transfer rate between
the directly excited high frequency mode and the overbarrier excitation of RC mode.

For the description of the indirect excitation of the RC mode via anharmonic couplings
with the high-frequency C--H(D) stretch mode one needs to include the anharmonic couplings
between the vibrational modes in  $H_{\mathrm{ph}}$, the vibrational part of the Hamiltonian.
In the lowest order only the cubic couplings have to be considered
\begin{multline}
\label{eq:Hph}
H_{\mathrm{ph}} =
\sum_{\nu}\hbar\Omega_{\nu} b_{\nu}^{\dagger} b_{\nu}\\
+\frac{1}{6}\sum_{\nu,\nu',\nu''}
\mathcal{K}_{\nu,\nu',\nu''}\left( b_\nu^{\dagger}+b_\nu \right)
\left( b_{\nu'}^{\dagger}+b_{\nu'}\right) \left( b_{\nu''}^{\dagger}+b_{\nu''}\right),
\end{multline}
where $\mathcal{K}_{\nu,\nu',\nu''}$ is the anharmonic coupling constant between
the vibrational modes $\nu$, $\nu'$, and $\nu''$. Intuitively it is clear that
we are interested only in few addends in the last summation, those which
connect the  high frequency vibrational modes and the RC mode.
In Sec.~\ref{sec:Anharmonic_coeff} we provide the full analysis
of the anharmonic couplings between the vibrational modes.

\section{Vibrational modes of acetylene on Cu(001)}
\label{sec:DFT}

\begin{figure}[b]
\begin{centering}
\includegraphics[width=0.8\columnwidth]{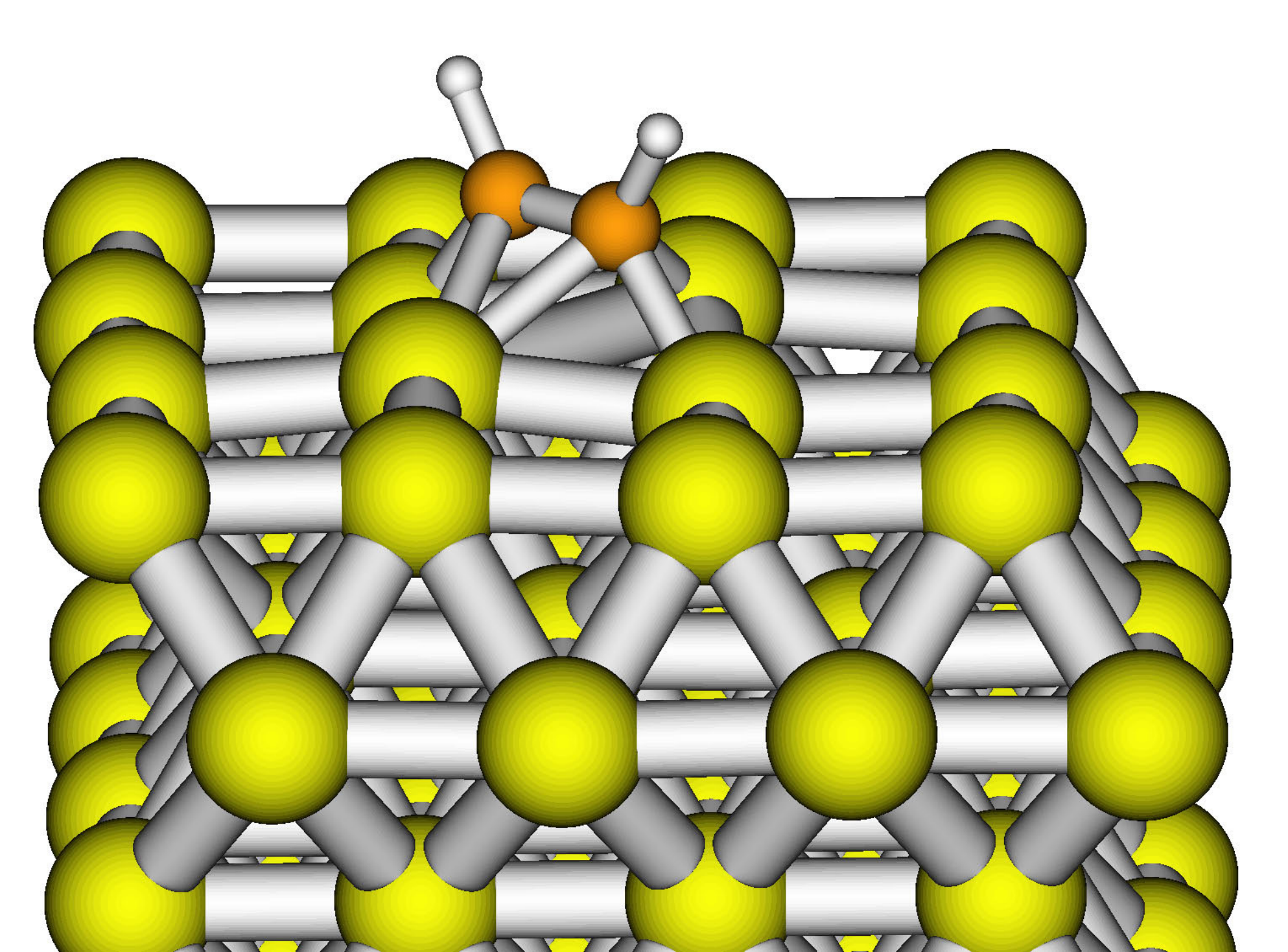}    %{FigS_1.eps}
\par
\end{centering}
\caption{\label{fig:Structure} Calculated equilibrium configuration of acetylene molecule (C$_2$H$_2$) on Cu(001).
The C--C and C--H bond lengths and C--C--H bond angle are given in the text.}
\end{figure}

The analysis of the  vibrational modes of a single C$_2$H(D)$_2$/Cu(001) is made by means of
DFT calculations on a 4$\times$4 Cu(001) surface with one
adsorbed acetylene molecule.\footnote{Calculational details include a real space cutoff of 200 Ry,
Gamma points approximation, double/single-z polarized (DZP/SZP) basis set for the C (DZP), H(DZP), and Cu(SZP) atoms.}
The calculations of relaxed geometries, vibrational energies, and
electron-hole pair damping rates are carried out with SIESTA.\cite{Soler2002,Frederiksen2007}

Figure~\ref{fig:Structure} shows schematically the equilibrium configuration of C$_2$H$_2$ on Cu(001) and the relaxation of the Cu atoms.
The calculated equilibrium configuration of the chemisorbed C$_2$H$_2$ molecule on Cu(001) is in agreement with
Ref.~\onlinecite{Olsson2002}: $d_{\rm CC}=1.40$~\AA, $d_{\rm CH}=1.12$~\AA ~and C--C--H bond angle is $120.7^\circ$.

The results  for  the vibrational
energies $\hbar \Omega_\nu$ and  electron-hole damping rates $\gamma_{\mathrm{eh}}^{(\nu)}$, $\nu = 1, \ldots 12$
are given in Tab.~\ref{tab:VibModes},
in comparison with the previous theoretical~\cite{Olsson2002} and experimental~\cite{Marinova1987,Avery1985,Stipe1998a} results.
Due to the lack of the experimental data available for the vibrational modes of the deuterated acetylene on Cu(001) surface
we provide the comparison with the vibrational modes for the C$_2$D$_2$ on Cu(110).\cite{Avery1985}
For C$_2$H$_2$ molecule the vibrational frequencies on Cu(110) and Cu(001) surfaces are similar.\cite{Avery1985,Marinova1987}

The last column of Tab.~\ref{tab:VibModes} gives (in relative units) the components
of the vibrational mode angular momenta
$\vec{L}^{(\nu)} = \sum_i \Omega_\nu (\vec{r}_i - \vec{r}_\mathrm{C.M.}) \times m_i \delta\vec{r}_i^{(\nu)}$,
where $\vec{r}_i$, $m_i$ are the  atomic positions (in the acetylene molecule) and masses, $\vec{r}_\mathrm{C.M.}$ the center-of-mass,
and $\delta\vec{r}_i^{(\nu)}$ the atomic displacements in the vibrational mode $\nu$.
Two vibrational modes No. 6 and No. 11  have non-zero z-component of the angular momentum and
are considered to be precursors of the rotational movement of the acetylene molecule on Cu(001)
in Ref.~\onlinecite{Stipe1998a}. Recently we have shown~\cite{Shchadilova2013a} that according to the analysis of the anharmonic couplings
and the ratio between the rotational barrier height and the energy of the vibrational mode
it is only possible that hindered rotational mode No. 6 is responsible for the signal observed in the experiment.~\cite{Stipe1998a}

%\paragraph{Barrier discussion and the ration between barrier height and the vibrational frequency}

The thermal activation energies  $\epsilon_B$ for the rotational mode of the C$_2$H$_2$
and C$_2$D$_2$ molecules on Cu(001)
are determined experimentally~\cite{Lauhon1999a} as $169\pm 3$~meV and $168\pm 4$~meV correspondingly.
Our estimate of  $\epsilon_B$ with the nudged elastic band method for C$_2$H$_2$/Cu(001) is $100$~meV,
which is only in a qualitative agreement with the experimental data.
The ratio of energy barrier height to the rotational mode energy $\Omega_r=78(63)$~meV
(we use the  experimental values\cite{Marinova1987,Avery1985})
is $\varepsilon_B / \Omega_r \approx 2.2 (2.6)$.

We also calculate by DFT the  transmission coefficients $\mathcal{T}^{(\nu)}$ for inelastic
electron to excite the vibrational modes
of both isotopologues of the acetylene on Cu(001). The electrons from the STM-tip
dominantly excite the C--H(D) vibrational stretch mode, the relevant transmission coefficients corresponding
to the symmetric and asymmetric modes are
$\mathcal{T}^{(1)}=1.3\times10^{11}$~(s$\times$V)$^{-1}$ and
$\mathcal{T}^{(2)}=5.9\times10^{11}$~(s$\times$V)$^{-1}$.
From the transmission coefficients we estimate the
probability factors to excite symmetric and asymmetric C--H(D) stretch modes as
$\zeta=\mathcal{T}^{(1)}\left(\mathcal{T}^{(1)}+\mathcal{T}^{(2)}\right)^{-1}=0.18$
and $(1-\zeta)=0.82$.

%For a more accurate estimation of the proportionality coefficient
%between the second power inelastic tunneling current and rotation rate
%we need to take into account the fact that the inelastic tunneling current is a sum of several components which
%arise from the scattering of tunneling elections on all relevant vibrational modes.
%In our case we are interested in two C--H(D) stretch modes, symmetric No. 1 and asymmetric No. 2.
%According to our DFT calculation the inelastic transmission through these modes are
%$\mathcal{T}^{(1)}=1.3\times10^{11}$~(s$\times$V)$^{-1}$ and
%$\mathcal{T}^{(2)}=5.9\times10^{11}$~(s$\times$V)$^{-1}$.
%In what follows we will take this into account. introducing the  probability factors to excite the symmetric C--H(D)
%stretch mode $\zeta=\mathcal{T}^{(1)}\left(\mathcal{T}^{(1)}+\mathcal{T}^{(2)}\right)^{-1}=0.18$
%and the asymmetric one $(1-\zeta)=0.82$.

\section{Estimation of the anharmonic coefficients for the acetylene molecule on Cu(001)}
\label{sec:Anharmonic_coeff}

Essential step in the analysis of the elementary processes in the C$_2$H$_2$/D$_2$ molecule on Cu(001) surface is
the analysis of the anharmonic coupling $\mathcal{K}_{\nu,\nu',\nu''}$ between the vibrational modes
$\nu$, $\nu'$ and $\nu''$ of the adsorbed molecule. For this estimation we propose a simple model of the potential energy surface.
We describe the C--H/D, C--C bonds by springs on rods (the latter
to fix the central character of the forces):
\begin{equation}
U_\mathrm{spr}(r_{i},r_{j})=\frac{\mu_{ij} \omega_{ij}^{2}}{2}\left(\sqrt{||r_{i}-r_{j}||^{2}}-L_{ij}\right)^{2},
\end{equation}
where $\omega_{ij}$ and $L_{ij}$ are the spring parameters. The Lennard-Jones potential is used
to describe the interaction between the carbon atoms and nearest neighbor Cu as well as between
the  H(D) atoms and the nearest neighbor Cu,
\begin{equation}
U_\mathrm{LJ}(r_{i},r_{j})=\varepsilon_{ij}\left( \left(\frac{a_{ij}}{||r_{i}-r_{j}||}\right)^{12}-2 \left(\frac{a_{ij}}{||r_{i}-r_{j}||}\right)^{6}\right),
\end{equation}
where $\varepsilon_{ij}$ is the depth of the potential and $a_{ij}$ the position of its minimum. The parameters are chosen to reproduce
the vibrational frequencies and eigenvectors and are given in Tab.~\ref{tab:ToyModelParam}.

In order to estimate the anharmonic couplings between different modes we expand the potential energy to the third order
in $\delta\vec{r_{i}}$, the atomic displacements,
\begin{multline}
U(\lbrace\delta\vec{r_{i}}\rbrace)=
U_{0}+\frac{1}{2}\sum_{i,j}^{N}\sum_{\alpha,\beta=1}^{3}a_{ij}^{\alpha\beta}\delta r_{i}^{\alpha}\delta r_{j}^{\beta}+\\
+\frac{1}{6}\sum_{i,j,k}^{N}\sum_{\alpha,\beta,\gamma=1}^{3}b_{ijk}^{\alpha\beta\gamma}
\delta r_{i}^{\alpha}\delta r_{j}^{\beta}\delta r_{k}^{\gamma}+o\left(\left\Vert \delta r\right\Vert ^{3}\right).
\end{multline}
After rotating  to the basis of normal coordinates we obtain
\begin{multline}
U(\lbrace\delta\vec{\tilde{q}}_{i}\rbrace)=U_{0}+\frac{1}{2}\sum_{\nu=1}^{3N}\Omega_{\nu}^{2}\delta \tilde{q}_{\nu}\delta \tilde{q}_{\nu}+\\
+\frac{1}{6}\sum_{\nu,\nu',\nu''=1}^{3N}\mathcal{K}^{(c)}_{\nu,\nu',\nu''}
\delta \tilde{q}_{\nu}\delta \tilde{q}_{\nu'}\delta \tilde{q}_{\nu''}+o\left(\left\Vert \delta \tilde{q}\right\Vert ^{3}\right).
\label{Eq:3rd}
\end{multline}
where $\lbrace  \delta\vec{\tilde{q}}_{i} \rbrace$ is a set of normal coordinates
and $\mathcal{K}^{(c)}$ is classical anharmonic coupling between the vibrational modes of the system.
The transformation  from the original Cartesian coordinates to the  normal coordinates basis has a form
\begin{equation}
\mathcal{K}_{\nu,\nu',\nu''}^{(c)}=\sum_{i,j,k}^{N}\sum_{\alpha,\beta,\gamma=1}^{3}
b_{ijk}^{\alpha\beta\gamma}\frac{e_{\nu,i}^{\alpha}}{\sqrt{m_i}}\frac{e_{\nu',j}^{\beta}}{\sqrt{m_j}}\frac{e_{\nu'',k}^{\gamma}}{\sqrt{m_k}},
\end{equation}
where $e_{\nu,i}^{\alpha}$ is the eigenvector coefficient between the normal coordinate $\delta \tilde{q}_\nu$
and the shift of $i$-th atom in $\alpha$ direction
$\delta r^{\alpha}_i$, $\delta \tilde{q}_\nu = \sum_{i,\alpha} e^{\alpha}_{\nu,i}\delta r^{\alpha}_i$.

We introduce a dimensionless displacement vector $\delta q_{\nu}$ in order to quantize the vibrational modes,
so that $\delta \tilde{q}_{\nu}=\delta q_{\nu}\sqrt{\hbar (2\omega_{\nu})^{-1}}$
and arrive from coordinates to operators in a canonical way $\delta q_{\nu}\rightarrow \delta \hat{q}_{\nu}$. Then,
creation $b^{\dagger}_\nu$ and annihilation $b_\nu$ operators of the corresponding mode can be introduced
and the vibrational Hamiltonian is transformed to~(\ref{eq:Hph})
where the anharmonic coefficient $\mathcal{K }_{\nu,\nu',\nu''}$ is connected with
the anharmonic coefficient before quantization as
$\mathcal{K }_{\nu,\nu',\nu''} = \mathcal{K}^{c}_{\nu,\nu',\nu''}\hbar^{3/2}\left(2\sqrt{2\omega_{\nu}\omega_{\nu'}\omega_{\nu''}}\right)^{-1}$.

\begin{figure}[t]
\begin{centering}
\includegraphics[width=0.98\columnwidth]{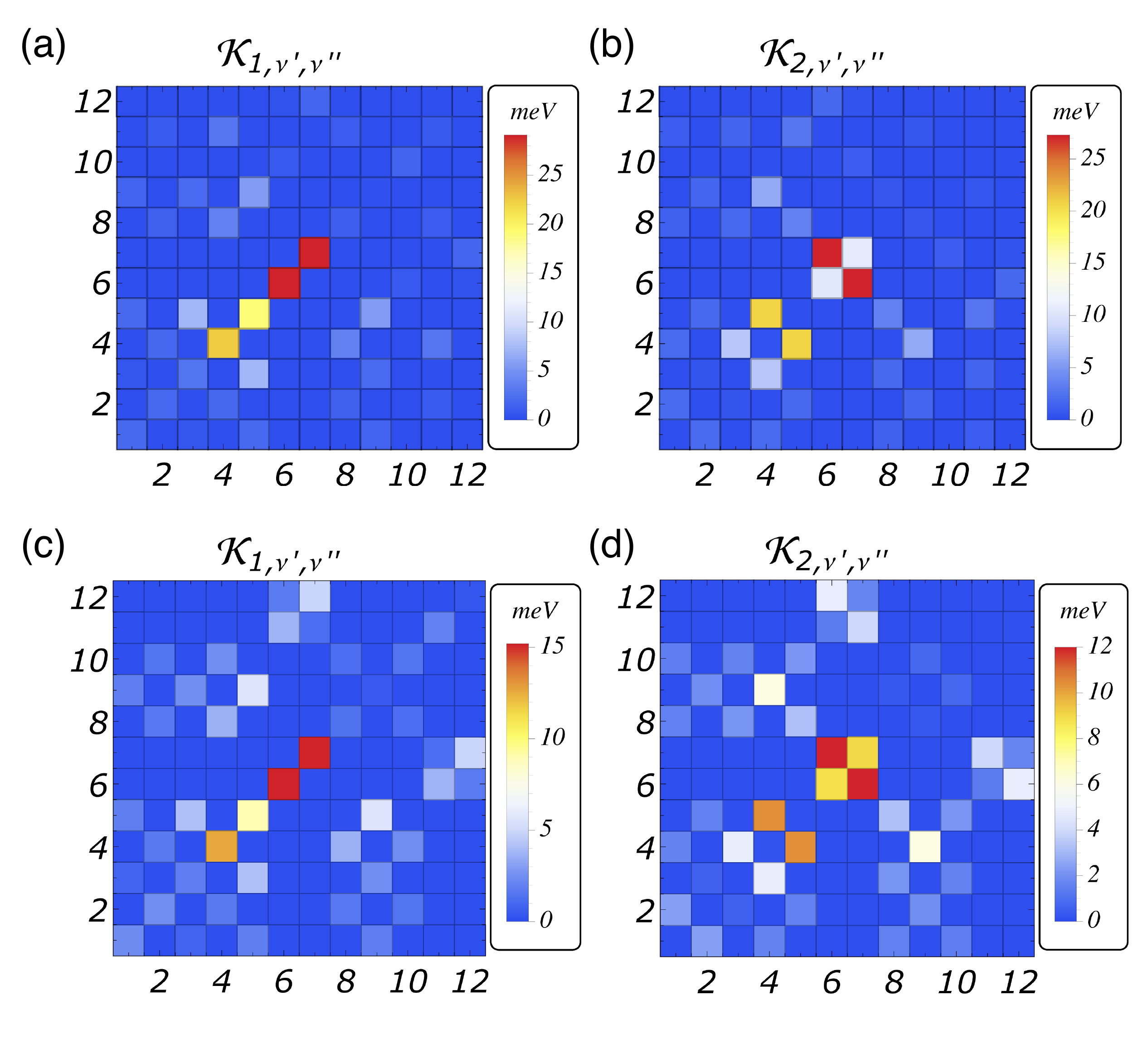}     %{Anh_all.eps}
\par
\end{centering}
\caption{\label{fig:Anh_coeff} The calculated anharmonic coefficients $\mathcal{K}_{\nu,\nu',\nu''}$
(a and b) of the C$_2$H$_2$/Cu(001), $\nu=1$ and $\nu=2$ respectively;
(c and d) of the deuterated acetylene C$_2$D$_2$ /Cu(001), $\nu=1$ and $\nu=2$ respectively.}
\end{figure}

The resulting anharmonic coefficients $\mathcal{K}_{\nu,\nu',\nu''}$ for the C--H(D) symmetric stretch mode $\nu=1$
and the C--H(D) asymmetric stretch mode $\nu=2$ are shown in Fig.~\ref{fig:Anh_coeff}.
These modes are known to be responsible for the high-energy threshold $\sim 360~(275)$~meV,
they are excited directly by the tunneling electrons.
The patterns of the anharmonic coefficients are similar for both isotopologues.
The absolute values of the anharmonic coefficients of the vibrational modes of the C$_2$D$_2$/Cu(001)
are twice lower than in case of C$_2$H$_2$/Cu(001).

The symmetric C--H(D) stretch mode No. 1 decays most efficiently via excitation of a pair of equivalent phonons:
No. 4 C--H(D) asymmetric in-plane bend or wag,
No. 5 in-plane bend or scissor, No. 6 out-of-plane bend or  asymmetric rotation, No. 7 cartwheel.
Coupling of the symmetric C--H(D) stretch mode No. 1 with the symmetric rotation mode No. 11 is ineffective since the corresponding anharmonic
coefficient is two orders of magnitude smaller than that for a coupling with the pair of out-of-plane bend or asymmetric rotation mode No. 6.

The asymmetric C--H(D) stretch mode No. 2 decays most efficiently via excitation of a pair of non-equivalent phonons, e.g.,
the pair of the asymmetric rotation No. 6 and cartwheel mode No. 7.

This simple estimation of the anharmonic coefficient shows that the rotation of both isotopologues of the acetylene molecule
is initiated via excitation of the asymmetric rotation mode No. 6.
There are two processes leading to the excitation of the reaction coordinate mode No. 6, the excitation
of a pair of the asymmetric rotation phonons  or the excitation of one asymmetric rotation phonon and one
phonon of the cartwheel mode No. 7.

\section{Anharmonic processes calculation. Keldysh formalism}
\label{sec:Keldysh_formalism}

We have shown in Sec.~\ref{sec:Anharmonic_coeff} that the excitation process
of the vibrational mode of the acetylene molecule  involves two possible pathways,
via double excitation of the reaction coordinate mode or
via excitation of the reaction coordinate mode and the auxiliary idler mode.
This allows us to write
the vibrational  Hamiltonian $H_{\mathrm{ph}}$ as
\begin{eqnarray}
\label{Eq:Hph}
H_{\mathrm{ph}} & = & H_0 + H_{\mathrm{ph},1} + H_{\mathrm{ph},2} \equiv
\sum_{\nu=h,r,i}\hbar\Omega_{\nu} b_{\nu}^{\dagger} b_{\nu} \\  \nonumber
&&
+\frac{1}{2}\mathcal{K}_{h,r,r}\left( b_r^{\dagger} b_r^{\dagger} b_{h} +\mathrm{h.c.} \right)
+ \mathcal{K}_{h,r,i}\left( b_r^{\dagger} b_i^{\dagger} b_{h} +\mathrm{h.c.}\right)
,
\end{eqnarray}
where  $\mathcal{K}_{h,r,r}$, $\mathcal{K}_{h,r,i}$ are the anharmonic coupling constants,
and $\nu =i$ is some auxiliary (idler) vibrational mode excited simultaneously
with the RC mode.

We discuss here both of the scenarios using the Keldysh diagram technique.\cite{Keldysh1964}
In both cases under consideration the frequencies of the vibrational modes are far from the resonance,
$\Omega_{h} > 2 \Omega_{r} \sim \Omega_r+\Omega_i$ and the anharmonic interaction between them can be treated as weak.

In what follows we derive the excitation rate of the RC mode due to the process described by $H_{\mathrm{ph},2}$;
the excitation rate due to $H_{\mathrm{ph},1}$ can be obtained replacing the index $i$ of the idler
phonon  in all formulas below, with the index  $r$ of the RC phonon.

For the description of the effective stationary occupation densities of the RC
mode we use the Keldysh-Green's
function method and  kinetic equation.\cite{Keldysh1964,Tikhodeev2004}
The anharmonic component of the excitation rate of the RC mode is given by the
one-loop polarization operator. Neglecting the temperature corrections, it reads
\begin{multline}
\label{Eq:Ga}
\Gamma_{\mathrm{in}}(\omega,\Omega_r,V)=
2\pi \mathcal{K}_{h,r,i}^2\int n_{\mathrm{ph}}^{(h)}(\varepsilon+\omega)\rho_{\mathrm{ph}}^{(h)} (\varepsilon+\omega)\\
\left[ 1+ n_{\mathrm{ph}}^{(i)}(\varepsilon)\right] \rho_{\mathrm{ph}}^{(i)}(\varepsilon) d\varepsilon,
 \end{multline}
where $\rho_{\mathrm{ph}}^{(\nu)}(\varepsilon)$ is the density of states of the RC and high-frequency modes
and $n_{\mathrm{ph}}^{(\nu)}(\varepsilon)$ are the corresponding occupation densities.
Formula (\ref{Eq:Ga}) describes the energy transfer rate to the hindered rotation mode of an adsorbate due
to the anharmonic coupling with the C--H stretch mode.

We proceed with the calculation of a total
excitation rate $\Gamma_{\mathrm{iet}}(\Omega_r,V)$ of the RC mode
\begin{equation}
\label{Gtot}
\Gamma_{\mathrm{iet}}(\Omega_r,V)=\int \Gamma_{\mathrm{in}}(\omega,\Omega_r,V)\rho_{\mathrm{ph}}^{(r)}(\omega) d \omega.
\end{equation}
After making a substitution of (\ref{Eq:Ga}) into (\ref{Gtot}) and in the saddle-point approximation the total RC excitation rate takes the form
\begin{multline}\label{PhonGenRate}
\Gamma_{\mathrm{iet},2}(\Omega_r,V)=
2\pi \mathcal{K}^2_{h,r,i}  n_{\mathrm{ph}}^{(h)}(\Omega_i+\Omega_r) \rho_{\mathrm{ph}}^{(h)}(\Omega_i+\Omega_r) \\
+2\pi \mathcal{K}^2_{h,r,i}
n_{\mathrm{ph}}^{(h)}(\Omega_h) \left(\rho^{(r)}_\mathrm{ph}(\Omega_h-\Omega_i)+\rho_{\mathrm{ph}}^{(i)}(\Omega_h-\Omega_r)\right).
 \end{multline}
The second term in Eq.~(\ref{PhonGenRate}) shows a threshold dependence on bias voltage
with high threshold value $\Omega_h$,
because it is proportional to the high-frequency mode occupation numbers
$n_{\mathrm{ph}}^{(h)}(\Omega_{h}) = \Gamma_\mathrm{iet}(\Omega_{h})\left(2\gamma^{(h)}_{eh}(\Omega_{h})\right)^{-1}$.
It can be shown  that the first term in~(\ref{PhonGenRate}) can be omitted due to the fact that
$n_{\mathrm{ph}}^{(h)}(\Omega_i+\Omega_r) \ll n_{\mathrm{ph}}^{(h)}(\Omega_{h})$.

The total excitation rate of the RC phonons due to the anharmonic term $H_{\mathrm{ph},2}$ is
\begin{multline}\label{eq:PhonGenRate2}
 \Gamma_{\mathrm{iet},2}(\Omega_r,V)\approx  \\
 2\pi \mathcal{K}^2_{h,r,i}  \frac{\Gamma_{iet}(\Omega_{h})}{2\gamma^{(h)}_{eh}(\Omega_{h})}
 \left[\rho^{(r)}_\mathrm{ph}(\Omega_h-\Omega_i)+\rho_{\mathrm{ph}}^{(i)}(\Omega_h-\Omega_r)\right],
\end{multline}
and  that due to  the anharmonic term $H_{\mathrm{ph},1}$ is
\begin{equation}\label{eq:PhonGenRate3}
\Gamma_{\mathrm{iet},1}(\Omega_r,V) \approx
4\pi \mathcal{K}^2_{h,r,r} \frac{\Gamma_{iet}(\Omega_{h})}{2\gamma^{(h)}_{eh}(\Omega_{h})}\rho^{(r)}_\mathrm{ph}(\Omega_h-\Omega_r).
\end{equation}
Note that both rates are proportional to the C--H(D) stretch mode excitation rate $\Gamma_\mathrm{iet}(\Omega_{h})$
and to small phonon densities of the RC and idler vibrational modes far from the resonance.

\section{Ladder climbing with two- and one-step processes} \label{sec:Ladder}
\begin{figure}[t]
\begin{centering}
\includegraphics[width=0.99\columnwidth]{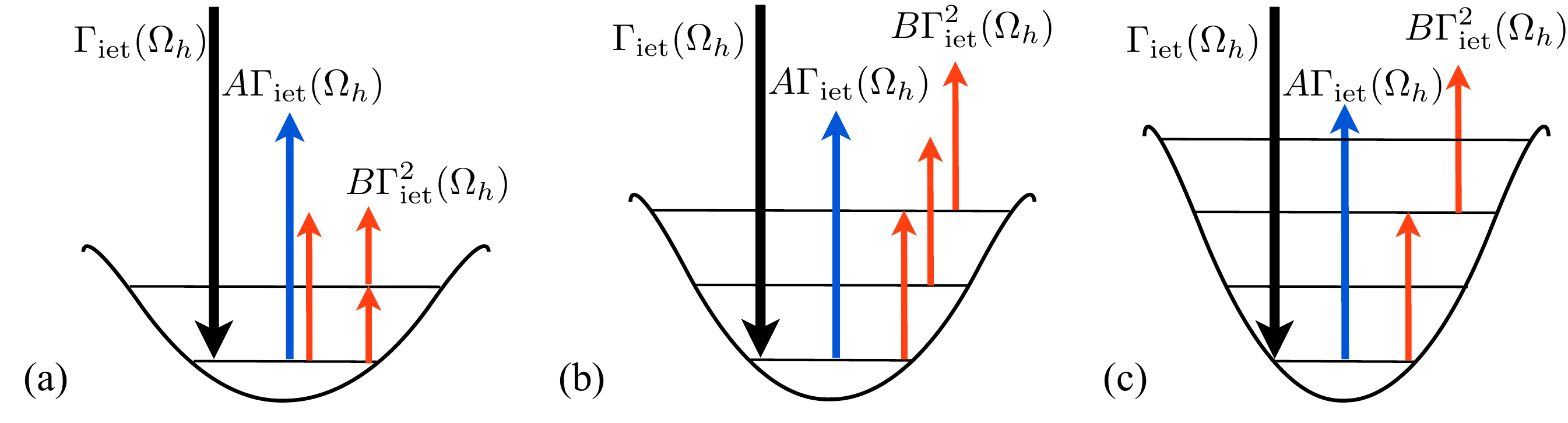}   %{Barriers_compiled.eps}
\par
\end{centering}
\caption{\label{fig:Barrier} Sketch of the linear and non-linear excitations of the RC mode. Three cases are considered:
(a) low RC barrier  $\Omega_r<\varepsilon_B<2\Omega_r$;
(b) intermediate RC barrier $2\Omega_r<\varepsilon_B<3\Omega_r$;
(c) high RC barrier $3\Omega_r<\varepsilon_B<4\Omega_r$.}
\end{figure}

Using the expression~(\ref{eq:PhonGenRate2}) or~(\ref{eq:PhonGenRate3})
and Pauli master equation we are able to calculate the rotational probability rate $R_B(V)$.
In this section we calculate the excitation rate of the RC phonons $R_{B}(V)$  using Pauli master equation.~\cite{Gao1997}
Several processes of
the rotation initiation of the acetylene molecule on Cu(001)
--- with low, intermediate, and high barrier ---  have  to be considered depending on the height of the rotational barrier, see in
Fig.~\ref{fig:Barrier}.
The situations with low and intermediate barrier height may take place
in case of C$_2$H$_2$ molecule on Cu(001). While in case of C$_2$D$_2$ the processes
 with high and intermediate barrier height  are  more probable, because of a lower energy
 of the hindered rotation mode No. 6 (see in Tab.~\ref{tab:VibModes}).

\subsection{Low RC  barrier}\label{sec:LowBarrier}
If the rotational barrier height is $\Omega_r<\varepsilon_B<2\Omega_r$, the one-step ladder climbing process
takes place.\cite{Gao1994} The excitation rate of this process $\Gamma_{\mathrm{iet},2}(\Omega_r,V)$ is given by (\ref{eq:PhonGenRate2})
and the relaxation rate is $\gamma_{\mathrm{eh}}^{(r)}$, thus the Pauli master equation can be written as
\begin{multline} \label{Eq:Pauli1step}
\frac{dP_{m}}{dt}  =  (m+1)\gamma_{\mathrm{eh}}^{(r)}P_{m+1}+m\Gamma_{\mathrm{iet},2}(\Omega_r,V)P_{m-1}\\
-  \left[m\gamma_{\mathrm{eh}}^{(r)}+(m+1)\Gamma_{\mathrm{iet},2}(\Omega_r,V)\right]P_{m}.
\end{multline}
The stationary solutions (in respect to $P_{0}$) for $m=1$ states in the localization potential of the RC mode
can be written as $P_{0}=1$, $P_{1}=\Gamma_{\mathrm{iet},2}(\Omega_r,V)\left(\gamma_{\mathrm{eh}}^{(r)}\right)^{-1}P_{0}\ll P_{0}$.

Reaction rate $R_B(V)$ is defined as a probability rate to overcome the localization potential barrier and in our notations
it is the excitation rate from the first excited level
\begin{equation}\label{eq:RB1}
R^{(1)}_B(V)=2\Gamma_{\mathrm{iet},2}^2(\Omega_r,V)\left(\gamma_{\mathrm{eh}}^{(r)}\right)^{-1}.
\end{equation}

Substituting the expression for the excitation rate of the RC phonons (\ref{eq:PhonGenRate2})
and (\ref{eq:PhonGenRate3}) into Eq.~(\ref{eq:RB1}) and using the expression
for the high-frequency phonons occupation densities
$n_{\mathrm{ph}}^{(h)}(\Omega_{h})=\Gamma_{\mathrm{iet}}(\Omega_{h},V)\left(2\gamma_\mathrm{eh}^{(h)}\right)^{-1}$
we obtain the proportionality coefficient $B^{(1)}$ between the excitation reaction rate  and the phonon generation rate
 $R^{(1)}_{B}(V)=B^{(1)}\Gamma_{\mathrm{iet},2}(\Omega_{h},V)$,
\begin{multline}\label{eq:B1}
B^{(1)}=2\pi^2 \mathcal{K}^4_{h,r,i} (1-\zeta)^2\frac{1}{\left(\gamma_\mathrm{eh}^{(h)}\right)^{2} \gamma_{\mathrm{eh}}^{(r)}}\\
\left(\rho^{(r)}_\mathrm{ph}(\Omega_h-\Omega_i)+\rho_{\mathrm{ph}}^{(i)}(\Omega_h-\Omega_r)\right)^2.
\end{multline}
where $h$, $r$, and $i$ correspond to the vibrational modes No. 2,  6, and  7, respectively.

Moreover, in  the considered here case of a  lower reaction barrier, $\Omega_r<\varepsilon_B<2\Omega_r$, the process of double excitation of
the reaction coordinate phonons  gives a contribution to the linear part $R_A(V)$ of the rotation probability.
The rotation rate can be estimated then as $R_A(V)\approx \Gamma_{\mathrm{iet},1}$.
Then the impact of this process into the proportionality coefficient $A$ can be written as
\begin{equation}\label{eq:A1}
A^{(1)}= 4\pi \mathcal{K}^2_{h,r,r} \zeta \tau^{(h)} \rho^{(r)}_\mathrm{ph}(\Omega_h-\Omega_r).
\end{equation}
where $h$ and $r$ correspond to the vibrational modes No. 1 and 6, respectively.

\subsection{Intermediate RC  barrier}\label{sec:IntermBarrier}
If the rotational barrier height is $2\Omega_r<\varepsilon_B<3\Omega_r$,  the excitation rate is assumed to be dominated by
the two-phonon anharmonic coupling with the high-frequency mode.
The pair of RC phonons excitation rate $\Gamma_{\mathrm{iet},1}(\Omega_r,V)$
is given by (\ref{eq:PhonGenRate3})
and the de-excitation process is dominated by the single phonon relaxation rate
$\gamma_{\mathrm{eh}}^{(r)}$. Then, according to Ref.~\onlinecite{Gao1994}, the Pauli master equation takes the  form
\begin{multline}\label{Eq:Pauli2step}
\frac{dP_{m}}{dt}=(m+1)\gamma_{\mathrm{eh}}^{(r)}P_{m+1} +m(m-1)\Gamma_{\mathrm{iet},1}(\Omega_r,V)P_{m-2}\\
-\left[m\gamma_{\mathrm{eh}}^{(r)}+(m+2)(m+1)\Gamma_{\mathrm{iet},1}(\Omega_r,V)\right]P_{m}.
\end{multline}
The stationary solutions (in respect to $P_{0}$) for $m=0,1,2$ states in the localization potential of the RC mode
can be written as $P_{0}=1$, $P_{1}=2\Gamma_{\mathrm{iet},1}(\Omega_r,V)\left(\gamma_{\mathrm{eh}}^{(r)}\right)^{-1}P_{0}\ll P_{0}$ and
\begin{equation}
P_{2}=\frac{\left(\gamma_{\mathrm{eh}}^{(r)} + 6\Gamma_{\mathrm{iet},1}(\Omega_r,V)\right)}
{2\gamma_{\mathrm{eh}}^{(r)}}P_{1}\approx\frac{\Gamma_{\mathrm{iet},1}(\Omega_r,V)}{\gamma_{\mathrm{eh}}^{(r)}}P_{0}.
\end{equation}

Reaction rate $R_B(V)$ in this case is a sum of the excitation rates from the first excited state
$6\Gamma_{\mathrm{iet},1}(\Omega_r,V)P_1$ and from  the second excited state $12\Gamma_{\mathrm{iet},1}(\Omega_r,V)P_{2}$,
\begin{equation}\label{eq:RB2}
R^{(2)}_B(V)=24\Gamma_{\mathrm{iet},1}^2(\Omega_r,V)\left(\gamma_{\mathrm{eh}}^{(r)}\right)^{-1}.
\end{equation}

Analogously, the coefficient $B^{(2)}$ between $R^{(2)}_{B}(V)$ and $\Gamma_{\mathrm{iet},1}^2(\Omega_{h},V)$ becomes
\begin{equation}\label{eq:B2}
B^{(2)}=48\pi^2 \mathcal{K}^4_{h,r,r} \zeta^2 \frac{\left(\rho^{(r)}_\mathrm{ph}(\Omega_h-\Omega_r)\right)^2}
{\left(\gamma_\mathrm{eh}^{(h)}\right)^{2}\gamma_{\mathrm{eh}}^{(r)}}.
\end{equation}
where $h$ and $r$ correspond to the vibrational modes No. 1 and 6,  respectively.

\subsection{High RC barrier}\label{sec:HighBarrier}
If the height of the RC barrier is even higher, $3\Omega_r<\varepsilon_B<4\Omega_r$,  the contribution
only from the second excited state should be considered and the final value of the
rotation probability is twice lower than in case of $2\Omega_r<\varepsilon_B<3\Omega_r$,
\begin{equation}\label{eq:RB2}
R^{(3)}_B(V)=12\Gamma_{\mathrm{iet},2}^2(\Omega_r,V)\left(\gamma_{\mathrm{eh}}^{(r)}\right)^{-1}.
\end{equation}

The coefficient $B^{(3)}$ between $R^{(3)}_{B}(V)$ and $\Gamma_{\mathrm{iet},1}^2(\Omega_{h},V)$,
\begin{equation}\label{eq:B3}
B^{(3)}=\frac{1}{2} B^{(2)} .
\end{equation}

\section{Fit of the rotation rate as a function of  bias voltage and anharmonic currents estimation}
\label{sec:Discussion}

In Sec.~\ref{sec:Ladder} we showed that the reaction rate
in all cases is a quadratic function of the RC phonon excitation rate which is a feature of the two-step ladder climbing process
and differs only in a proportionality coefficient
\begin{equation}
R_B(V)=B \Gamma_\mathrm{iet}^2(\Omega_{h},V).
\end{equation}

To reproduce the experimental results from Ref.~\onlinecite{Stipe1998a}
we use the exact expressions for the total current $I(V)$ and the
generation rate of the high-frequency phonons, see Ref.~\onlinecite{Tikhodeev2004}.
We tune the parameters of the model $\varepsilon_a(0)$, $\Delta_s$,  $\Delta_t$
to fix the value of the total tunneling current, the lifetime of the high-frequency vibrational mode
due to electron-hole pairs excitations and the ratio between inelastic and total tunneling current.
Dependence of the results on the value of the adsorbate electron energy $\varepsilon_a(0)$
is weak assuming that the $\varepsilon_a(0)$ is far from the Fermi energy  $\varepsilon_F$,
$\mu_s-\varepsilon_a(0)\ll \Gamma_s+\Gamma_t$ . We fix the value to be $\varepsilon_a(0)=2$~eV.
According to our estimate this is in agreement with the asymmetric IETS signal
reported in Ref.~\onlinecite{Lauhon2001}.

\subsection{C$_2$H$_2$/Cu(001)}

\begin{figure}[t]
\begin{centering}
\includegraphics[width=1\columnwidth]{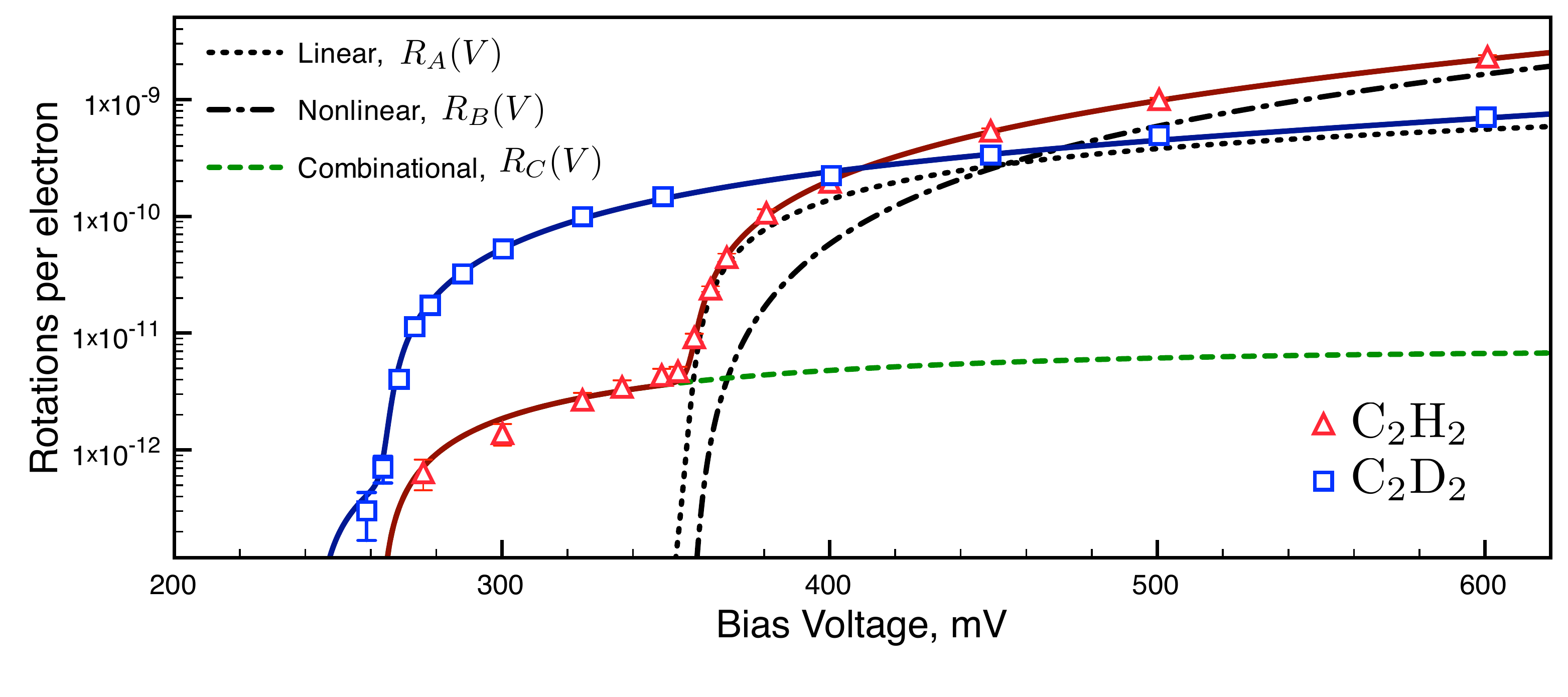}      %{RvV_01.eps}
\end{centering}
\caption{\label{fig:Rotation_rate_C2H2}
Comparison of the rotation yield per electron $Y$ as a function of bias voltage for two isotopologues of acetylene
C$_2$H$_2$ and C$_2$D$_2$ on Cu(001) surface.
Symbols are the experimental data from Ref.~\onlinecite{Stipe1998a}.
Partial processes $R_A(V)$, $R_B(V)$ and $R_C(V)$ of the rotation rate of the C$_2$H$_2$ molecule are shown with short dashed, dash-dotted and dashed lines, respectively.
Solid lines correspond to the sum of all partial processes.}

\end{figure}

\begin{figure}[t]
\begin{centering}
\includegraphics[width=1\columnwidth]{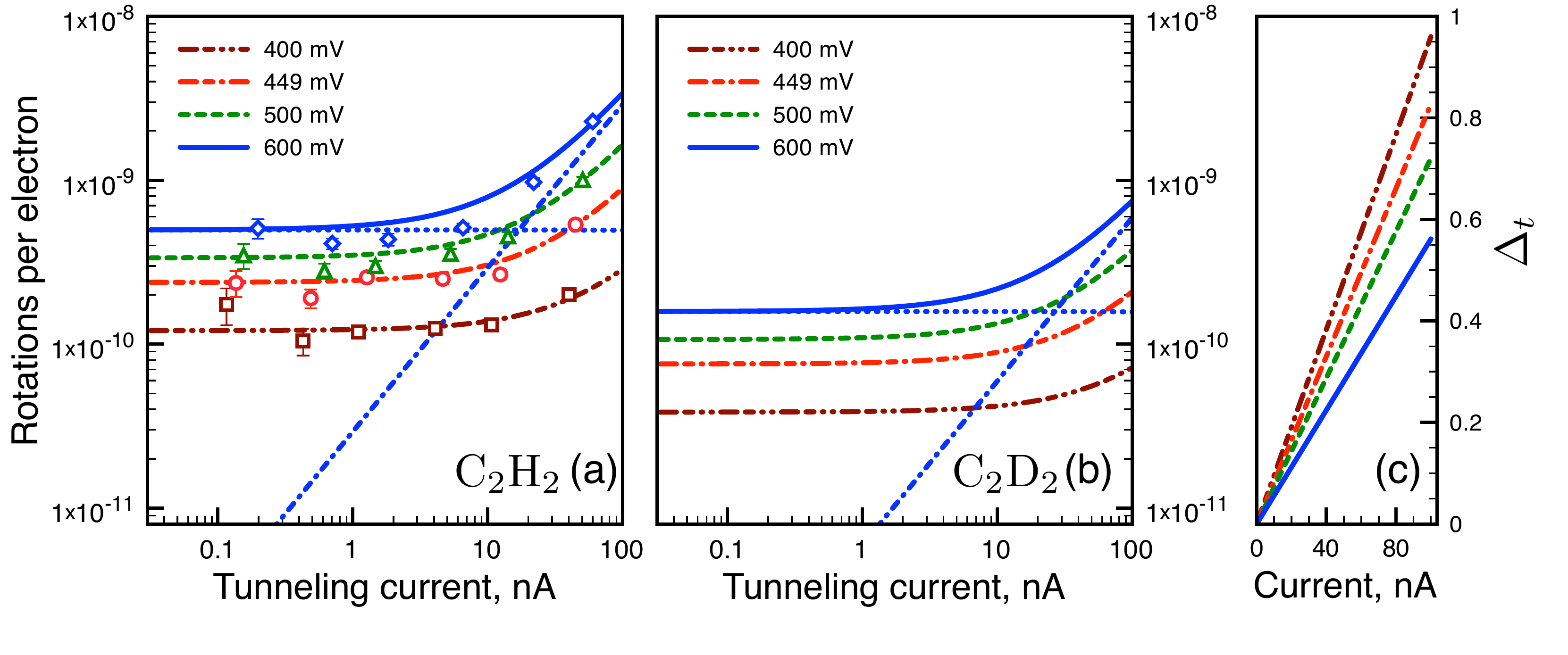}    %{RvC.eps}
\end{centering}
\caption{\label{fig:RvC}
(a)-(b) Rotation yield per electron $Y$ as a function of tunneling current for two isotopologues of acetylene
C$_2$H$_2$ and C$_2$D$_2$ on Cu(001) surface.
Lines are calculated results for different bias voltages: $V=400$, $449$, $500$, $600$~mV.
Thin dash-dotted lines are the linear and nonlinear contributions to the rotational
yield for $V=600$~mV. Dots in panel (a) correspond to the experimental data from Ref.~\onlinecite{Stipe1998a};
(c) The dependence of the molecule-tip hybridization parameter $\Delta_t$ on the tunneling current
for different bias voltages.}
\end{figure}

We fix the total current $I=40$~nA and the ratio of the inelastic component to the total value of the tunneling conductance
$\sigma_\mathrm{in}/\sigma \approx 0.1$ (this value follows from our DFT analysis). Then  we obtain the following parameters of
the hybridization between leads and the molecular orbital: $\Delta_s=250$~meV, $\Delta_t=9.6$~meV.
The main vibrational mode is the C--H stretch mode with the frequency $\Omega_h=358$~meV,
and the lifetime $\tau_\mathrm{ph}=1$~ps ($\gamma_\mathrm{eh}=1$~ps$^{-1}$) obtained from our DFT calculations.
These parameters allow us to calculate elastic and inelastic components of tunneling current as  functions of
the bias voltage and to fit the experimental data. The fitting coefficient of the linear contribution
is $A=3 \times 10^{-6}$ and of the quadratic contribution is $B=7\times10^{-16}$~s$^{-1}$.
In Fig.~\ref{fig:Rotation_rate_C2H2} the fit of the experimental data~\cite{Stipe1998a} of the
C$_2$H$_2$/Cu(001) rotational rate is shown as a function of bias voltage
for the tunneling current $I=40$~nA.
The dependence of the rotational rate as a function of the tunneling current is obtained with the
same parameters as described above, see in Fig.~\ref{fig:RvC}(a).
This dependence is calculated
by varying the molecule-tip hybridization parameter $\Delta_t$ as shown in Fig.~\ref{fig:Rotation_rate_C2H2}(c).

As we discussed in the previous section, there are  two possible ways of the RC excitations. Using the parameters above
and the estimate of the anharmonic coefficients $\mathcal{K}_{\nu,\nu',\nu''}$, the estimation for $B$
is made.

In case of a low energy barrier $\Omega_r<\varepsilon<2\Omega_r$ the expression for the coefficient $B$
is given by Eq.~(\ref{eq:B1}). We obtain $B^{(1)}=3\times10^{-15}$~s
which is approximately of the same order of magnitude as the value $B=7\times 10^{-16}$~s obtained from the best fit to the experimental data.
The full set of parameters used for approximation $B^{(1)}$ and $A^{(1)}$ for C$_2$H$_2$/Cu(001) is
$\mathcal{K}_{h,r,i}\approx 30$~meV,
$\gamma^{(a)} = 0.2$~ps$^{-1}$, $\gamma^{(h)}=0.7$~ps$^{-1}$,
$\gamma_r=0.7$~ps$^{-1}$, $\Delta=\Omega_h-\Omega_r-\Omega_i\approx 202$~meV ($\Omega_i\approx\Omega_r$).
In this case the contribution to the linear coefficient $A$ is given by Eq.~(\ref{eq:A1}). Using the same set of the parameters
we obtain the estimated value of $A^{(2)}=6\times 10^{-3}$
which is three orders of magnitude larger than the value of $A$ obtained from the fitting to the experimental data.
Evidently, this makes the case of a lower reaction coordinate
barrier $\Omega_r<\varepsilon_B<2\Omega_r$ hardly possible.

In the other case of intermediate rotational barrier $2\Omega_r<\varepsilon_B<3\Omega_r$
the expression for the coefficient $B$ is given by Eq.~(\ref{eq:B2}).
Using the parameters  for C$_2$H$_2$/Cu(001):
$\mathcal{K}_{h,r,r}\approx 30$~meV,
$\gamma^{(h)}_\mathrm{ph}=\gamma^{(1)}=1$~ps$^{-1}$, $\gamma^{(r)}=0.7$~ps$^{-1}$,
$\Delta=\Omega_h-2\Omega_r\approx 202$ meV,
we obtain $B^{(2)}\approx 2\times 10^{-15}$s,
which is the same order of magnitude as the best fitted parameter $B=7\times 10^{-16}$~s.

As we can see, our rough estimate of the anharmonic coefficients gives us nevertheless an opportunity to
obtain the amplitude of the nonlinear contribution  $B$ in the reaction yield quantitatively.
This is mostly due to a large amount of the experimental data available in Ref.~\onlinecite{Stipe1998a} which allows us
to fix all the parameters of our estimation.

\subsection{C$_2$D$_2$/Cu(001)}

As in case of C$_2$H$_2$/Cu(001), we fix the total current $I=40$~nA
and the ratio of the inelastic component to the total value of the tunneling conductance
$\sigma_\mathrm{in}/\sigma \approx 0.08$. This value corresponds to the following parameters of
the hybridization between leads and the molecular orbital: $\Delta_s=250$~meV, $\Delta_t=14$~meV.
The main vibrational mode is the C--D stretch mode with the frequency $\Omega_h=265$~meV,
and the lifetime $\tau_\mathrm{ph}=3.3$~ps ($\gamma_\mathrm{eh}=0.35$~ps$^{-1}$) obtained from our DFT calculations.

\begin{figure}[t]
\begin{centering}
\includegraphics[width=1\columnwidth]{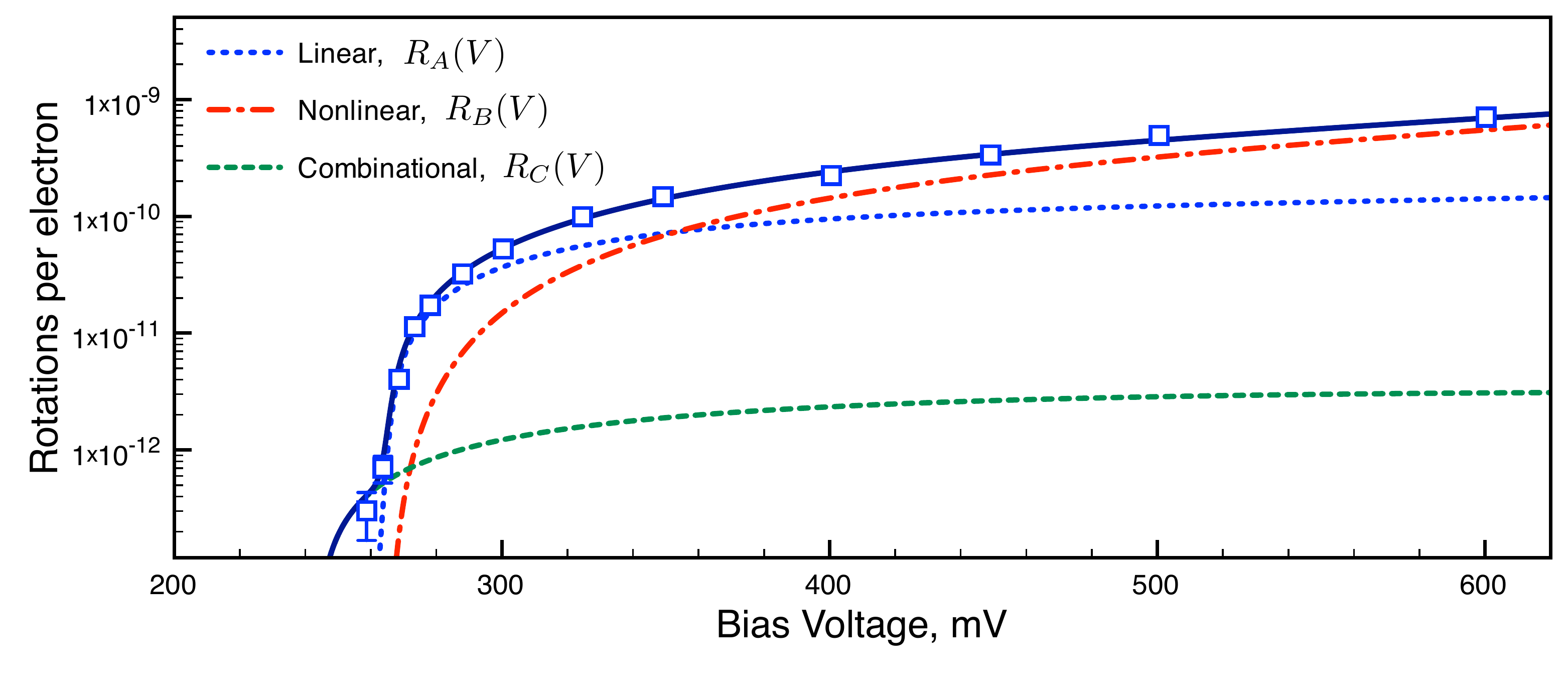}      %{RvV_C2D2.eps}%{figuresAnharmonicCoeff_v2.eps}
\par
\end{centering}
\caption{\label{fig:Rotation_rate_C2D2}Dependence of the rotation rate per electron of the C$_2$D$_2$ molecule on bias voltage.
Squares are the experimental data from Ref.~\onlinecite{Stipe1998a}.
The fitted rates $R_A(V)$, $R_B(V)$, and $R_C(V)$ are shown as short dashed, dash-dotted, and dashed
lines, respectively. The solid black line is the sum of all partial processes.}
\end{figure}

The fitting parameters for the one- and two-electron processes are equal to  $A=8 \cdot 10^{-7}$,
$B=2 \cdot 10^{-16}$ s. The fitted rates $R_A(V)$ and $R_B(V)$ are shown  as blue dashed and red dotted
curves in  Fig.~\ref{fig:Rotation_rate_C2D2}.
We also show the dependence of the rotation yield as a function of tunneling current for the
deuterated acetylene on Cu(001)  in Fig.~\ref{fig:RvC}(b).
As in case of
C$_2$H$_2$ molecule, we calculate the dependence on  tunneling current
by varying the molecule--tip hybridization parameter $\Delta_t$, as shown in Fig.~\ref{fig:Rotation_rate_C2H2}(c).
There is no experimental results available on the dependence of the rotation yield as a function of tunneling current
for the deuterated molecule to compare with. But the main feature
of the rotation yield of two isotopologues of the acetylene molecule on Cu(001) surface
--- the crossover from a single to two electron process --- is clearly distinguishable
for both isotopologues.

Since the experimental value of the rotational mode of the C$_2$D$_2$ molecule is unknown,
we consider two possible ways of the RC excitations for the theoretical estimation of
the coefficient $B$, the cases of intermediate~Eq.~\ref{eq:B2} and high~Eq.~\ref{eq:B3}  RC barrier.
The estimation gives us values $B^{(2)}=1\cdot 10^{-15}$~s and $B^{(3)}=5\cdot 10^{-16}$~s
correspondingly. The full set of parameters used for approximation of $B^{(2)}$ and $B^{(3)}$ for C$_2$D$_2$/Cu(001) is:
$\mathcal{K}_{h,r,r}=15$~meV,
$\Omega_h- 2\Omega_r\approx 166$~meV,
$\gamma^{(r)}_\mathrm{eh}=0.26$~ps$^{-1}$, $\gamma^{(h)}_\mathrm{eh}=0.35$~ps$^{-1}$,
$\zeta=\mathcal{T}^{(1)}/(\mathcal{T}^{(1)}+\mathcal{T}^{(2)})=0.2$.
This estimated coefficient is 5 or 2.5 times larger than that obtained from the fit to the experimental data, $B_\mathrm{fit}=2 \cdot 10^{-16}$~s.
We believe that both of them are in a reasonable agreement, bearing in mind the simplifications of the theoretical method.

\section{Combination band processes}
\label{sec:Combinational}

Considering the excitation of the acetylene isotopologues on Cu(001) we have to discuss also the excitation process of
the rotational motion below the main threshold. The magnitude of the reaction yield of
C$_2$H$_2$ molecule is very low but non zero and exhibits a lower threshold of $\sim 240-250$~mV.
The energy scale of this threshold is of the order of the vibrational energies but according to the vibrational modes
analysis in Tab.~\ref{tab:VibModes} there are no corresponding vibrational modes. Our proposal
is that several vibrational modes are involved in the electron-phonon scattering, i.e., the combination band process occurs.\cite{Shchadilova2013a}
Note that for the rotation rate of the deuterated acetylene molecule on Cu(001) surface
 no rotation below the high energy threshold at $275$~mV was observed.\cite{Stipe1998a}
To understand what happens here we consider the same type of the processes as in case of C$_2$H$_2$/Cu(001),
where the combination band process is well resolved in the experiment.

We consider the following process of an inelastic electron tunneling  which involves a
simultaneous combination band\cite{Jakob1998}  generation of two coherent
phonons $\nu = c1,c2$. Assuming that the adsorbate energy in Eq.~(\ref{eq:He}) is now a function
of these vibrational modes, $\varepsilon_{a}(\{ q_{c1},q_{c2}\})$ , and expanding it in a Taylor series
\begin{equation}\label{eq:Hcomb}
\varepsilon_{a}(\{ q_{c1},q_{c2}\})=\varepsilon_{a}(\lbrace 0 \rbrace)+\eta(b_{c1}^{\dagger}+b_{c1})(b_{c2}^{\dagger}+b_{c,2}),
\end{equation}
where $\eta=\partial^{2}\varepsilon_{a}(\lbrace 0\rbrace)/\left(\partial q_{c1}\partial q_{c2}\right)$ and,
$b_{c1}$ and $b_{c2}$ are the annihilation operators
of the vibrational modes with frequencies $\Omega_{c1}$ and $\Omega_{c2}$,  and damping rates
$\gamma^{(c1)}_\mathrm{eh}$, $\gamma^{(c2)}_\mathrm{eh}$.

In order to calculate the excitation rate of coherent phonons we use the Keldysh-Green's
function method.\cite{Keldysh1964,Tikhodeev2004}, neglecting the temperature corrections (i.e., assuming $T=0$).
It can be shown that  the form of a single-phonon process rate $\Gamma_{\mathrm{iet}}(\Omega,V)$,\cite{Tikhodeev2004}
Eq.~(\ref{Eq:Giet}) can be used, where the single vibrational frequency is replaced by  the sum of two vibrational frequencies.
This gives for the combinational reaction rate
\begin{equation}
\label{Eq:Rc}
R_C(V)=C \Gamma_\mathrm{iet}(\Omega_{c1}+\Omega_{c2},V),
\end{equation}
where $\Gamma_\mathrm{iet}(\Omega_{\mathrm{c},1}+\Omega_{\mathrm{c},2},V)=$ .
$$
\frac{\gamma^{(c1)}_\mathrm{eh}+\gamma^{(c2)}_\mathrm{eh}}{\Omega_{\mathrm{c},1}+\Omega_{\mathrm{c},2}}\frac{\Delta_t}{\Delta_{s}}
\left(\left|eV\right|-\Omega_{c1}-\Omega_{c2}\right) \Theta\left(\frac{\left|eV\right|}{\Omega_{c1}+\Omega_{c2}}-1\right).
$$
%and the coefficient $C$ depends on the parameters of the system.

\begingroup
\squeezetable
\begin{table}[t]
\caption{\label{tab:CohProcesses}
Possible coherent processes of the vibrational modes excitation in C$_2$H$_2$ and C$_2$D$_2$ molecules.
The frequencies and lifetimes of the vibrational modes are given in Tab.~\ref{tab:VibModes}.}
\begin{ruledtabular}
\begin{tabular}{l  l  l  l}
&& C$_2$H$_2$  & C$_2$D$_2$ \\
&& $\Omega_h=360$~meV &$\Omega_h=265$~meV \\
\hline
1 & No.  3 C--C stretch &$\sum \Omega_i=282$~meV & $\sum \Omega_i =241$~meV\\
& No. 5 C--H(D) in-plane bend  &-- & $C=5.1\cdot 10^{-9}$\\
& or scissor, symmetric & & \\
\hline
2 & No. 3 C--C stretch & $\sum \Omega_i =242$~meV & $\sum \Omega_i =243$~meV \\
& No. 6 C--H(D) asym rotation & $C=8.7\cdot 10^{-9}$ & $C=5.3\cdot 10^{-9}$\\
& or out-of-plane bend &&\\
\hline
3 &No. 4 C--H(D) in-plane bend & $\sum \Omega_i =248$~meV & $\sum \Omega_i =186$~meV\\
& or wag, asymmetric &$C=1.87\cdot 10^{-8}$ & $C=1.3\cdot 10^{-9}$ \\
& No. 5 C--H(D) in-plane bend &&\\
& or scissor, symmetric & & \\
\hline
4 &No. 4 C--H(D) in-plane bend& $\sum \Omega_i =220$~meV & $\sum \Omega_i =186$~meV\\
& or wag, asymmetric &$C=2.1\cdot 10^{-8}$ & $C=3.2\cdot 10^{-9}$ \\
& No. 6 C--H(D) asym rotation & &\\
& or out-of-plane bend &&\\
\end{tabular}
\end{ruledtabular}
\end{table}
\endgroup

Table~\ref{tab:CohProcesses} summarizes  the possible combinations of the vibrational modes
near the high-frequency threshold $\Omega_h$
for both acetylene isotopologues and
the coefficients C of the corresponding coherent processes.

For C$_2$H$_2$ molecule there are three possibilities (rows 2--4 in Tab.~\ref{tab:CohProcesses}) with similar  energies
which approximately correspond to the  experimentally observed lower  threshold $\sum \Omega_i\approx 240-250$~meV.
For C$_2$D$_2$ molecule there are four processes (rows 1--4 in Tab.~\ref{tab:CohProcesses})  with  two possible threshold energies,
$\sum \Omega_i \approx 185$~meV and $\sum \Omega_i \approx 240$~meV.
The coefficients $C$ estimated for rows 1 and 2  in case of C$_2$D$_2$
are of the same order of magnitude as for row 2 in case of C$_2$H$_2$.
But the threshold  energy of such combination band process in C$_2$D$_2$ is very close to
the high-frequency threshold  with the energy $\Omega_h\sim275$~meV.
The other two processes, rows 3 and 4,  are at least an order of magnitude weaker  in case of C$_2$D$_2$
than in case of C$_2$H$_2$.

Figures~\ref{fig:Rotation_rate_C2H2} and~\ref{fig:Rotation_rate_C2D2} show the combination band processes
for the C$_2$H$_2$ and C$_2$D$_2$ isotopologues respectively. We plot the rate of the
combinational process involving the vibrational modes No. 3 and No. 6 for the both isotopologues
(row 2 in Tab.~\ref{tab:CohProcesses}). The parameters  used in calculations of $C$  (the energies and lifetimes of the
vibrational modes) are given in Tab.~\ref{tab:VibModes}. The values $C$ obtained from the fit of the experimental
data  are $8.7\cdot 10^{-9}$ and $5.3\cdot 10^{-9}$ for C$_2$H$_2$ and C$_2$D$_2$ correspondingly.
Although it is difficult to make a theoretical estimate of this coefficient, the
fitting to the experiment shows that the
combination band single-electron process is about $A/C \sim 200$ times slower than the process
with rotation excitation via the C--H/D stretch mode. This is in a reasonable agreement with the fact that the process Eq.~(\ref{eq:Hcomb})
occurs in the next order of the perturbation theory  compared with the process Eq.~(\ref{eq:Ea}).

%The resulting rate for the coherent processes (1) and (2) is shown for
%deuterated acetylene in Fig.~\ref{fig:Rotation_rate_C2D2} as green dashed line.
%The black solid line is the sum of all process, it reproduces the experimental
%results for the deuterated acetylene quite well.
The fact that in the original experiment\cite{Stipe1998a} no signs of the combination band process
were observed in case of C$_2$D$_2$ thus finds
a simple explanation.  It appears that due to different
isotope shifts of different vibration modes (the isotope shift is maximal for
the C--H/D stretch modes), the threshold of the process via
the stretch mode excitation shifts in case of C$_2$D$_2$  to $\Omega_h =275$~ meV.
Whereas the combination band process threshold is only slightly shifted to  240 meV. As a result, the faster
C--D stretch mode process screens the combination band process in case of C$_2$D$_2$.

\section{Conclusions}
\label{sec:Conclusions}

We have carried out the thorough discussion of the excitation processes of the rotations of the acetylene
isotopologues on Cu(001) in STM-contact. Using the combination of the DFT calculations of the vibrational modes
of the adsorbed molecules and estimation of the coupling coefficients between the vibrational modes,
the reaction coordinate mode is identified. The linear and nonlinear processes thus are distinguished.
For the description of the nonlinear RC excitation process we apply the Keldysh diagram technique
for the nonequilibrium processes and the Pauli master equation for the stationary reaction rate calculation.
We analyse several scenarios depending on the height of the RC barrier and provide the comparison
of the experimentally obtained data for the acetylene isotopologues rotation on Cu(001) with the analytical
dependencies of the reaction yield as a function of bias voltage and the tunneling current.

We also discuss the possible processes of the excitation rotational motion of the acetylene molecule below
the main threshold $\sim \Omega_h$. We show that the combination band processes are likely given by the
combination of the vibrational modes that are only slightly shifted due to the isotope-effect in comparison
with the isotope-shift of the main threshold.

\begin{acknowledgments}
This work was supported in part by
the Federal Target Program ``Scientific and scientific-pedagogical personnel of innovative Russia" in 2009-2013, 
and the Presidential Grant for Leading Russian Science Schools (Prof. L.V. Keldysh school grant NSh-4375.2012.2.)
H.U. was supported by a Grant-in-Aid for Scientific
Research (Grants No. S-21225001 and No. B-1834008) from
Japan Society for the Promotion of Science (JASP).
\end{acknowledgments}

\appendix

\section{Parameters for the potential surface fit}

Table~\ref{tab:ToyModelParam} specifies the parameters used in the simple ``springs-on-rods'' model
for both C$_2$H$_2$ and C$_2$D$_2$ molecules on Cu(001). The parameters are obtained using random walk
method. Two criteria are introduced. The first one is an average relative error between
the eigenvalues of the simpler model and those obtained with DFT,
$\left( \overline{\omega^\mathrm{\; model}/\omega^\mathrm{\; DFT}}\right)$
this gives us some value in the range $\left[ 0, 1\right]$.
The other one is the cosine of the average angle between eigenvectors of the simpler model
and the eigenvectors obtained with DFT,
$\cos\left( \overline{\vec{e}_m^\mathrm{\; model}\cdot \vec{e}_m^\mathrm{\; DFT}}\right)$,  which is also in the range $\left[ 0, 1\right]$.
The product of these criteria is a controlled parameter we used to find the best fit values.
We restricted the number of tries to $10^6$ and the best
values of the criterion for C$_2$H(D)$_2$ on Cu(001) were 0.75 (0.83).

The comparison between the
frequencies of the vibrational modes obtained with DFT calculations and with simple model are given in Tab.~\ref{tab:VibModes2}.

%--------------------------------------------------------------------------------------------------------
%Table with the parameters of the spring on rod model
%--------------------------------------------------------------------------------------------------------
\begingroup
\squeezetable
\begin{table}[h]
\caption{\label{tab:ToyModelParam}
The parameters of the "springs-on-rods" model to reproduce the calculated by DFT vibrational modes  of C$_2$H(D)$_2$ on Cu(001)}
\begin{ruledtabular}
\begin{tabular}{c c c c c c c}									
Bonds  								& C--C 				&	C--H(D) 		& C--Cu  			& C--Cu					& H(D)-Cu  		& H(D)-Cu			   \\
										&						&						&(nearest)  		& (next-nearest)	&(nearest)  		& (next-nearest)	\\	
\hline	
$\omega$,~meV				& 161 (137)  	& 363 (262)  	& - 					& -  						& - 					& -\\
$L$, \AA  						& 1.38 (1.38)	& 1.04 (1.05)	& - 					& -  						& - 					& - \\
$\varepsilon_{ij}$,~meV 	&  - 					& - 					& 578 (556) 		& 86 (36) 				& 153 (248) 		& 55 (202) \\
$a_{ij}$, \AA  					& - 					& - 					& 1.94(1.94)		& 2.20 (2.21) 		& 2.55 (2.55)	& 3.05 (3.05) \\
\end{tabular}
\end{ruledtabular}
\end{table}
\endgroup

\begingroup
\squeezetable
\begin{table}[h]
\caption{\label{tab:VibModes2}
Vibrational eigenfrequencies $\hbar \Omega_\nu$ (in meV)
of C$_2$H$_2$ and C$_2$D$_2$  on Cu(110)  (DFT and simple model results)}
\begin{ruledtabular}
\begin{tabular}{l c c c c c c c c c c c c}
%$\nu$ & 1. $\nu_s$(CD)            & 2.  $\nu_{as}$(CD)           & 3. $\nu$(CC)           & 4. $\delta_{as}$(CD)           & 5. $\delta_{s}$(CD)          & 6. $\gamma$(CD)           &  7.       &  8. Cu-acetylene        &  9.  Cu-acetylene       &  10    & 11. Acetylene rotation     & 12.  \\
 $\nu$ & 1 & 2&3&4&5&6&7&8&9&10&11&12\\
 \hline
DFT C$_2$H$_2$/Cu(001)		& 371 	& 368 	& 167 	& 131 	& 111 		& \bf{100} 	& 71 		& 58 		& 50 		& 29 		& \bf{28} 		& 23 \\
model C$_2$H$_2$/Cu(001) 	& 371 	& 370 	& 167	& 111	& 106		& \bf{100}	& 100		& 61			& 49			& 29			& \bf{24} 		& 23 \\
\hline
DFT C$_2$D$_2$/Cu(001)		& 275 	& 270 	& 164 	& 108 	& 79 		& \bf{77} 		& 52 		& 50 		& 49 		& 29 		& \bf{26} 		& 22 \\
model C$_2$D$_2$/Cu(001) 	& 275 	& 273 	& 143	& 83		& 78			& \bf{76}		& 76			& 59			& 49			& 26			& \bf{26} 		& 20\\
\end{tabular}
\end{ruledtabular}
\end{table}
\endgroup

%\bibliography{biblio_01}
%\end{document}


\end{document}